\documentclass[12pt,a4]{article}
\usepackage{amsthm,amsmath,latexsym,amssymb,amsfonts,amscd}
\usepackage{graphics,fancyhdr,array,stmaryrd,euscript}
\usepackage{slashed,tikz}
\numberwithin{equation}{section}

\usepackage{hyperref}

\usepackage[numbers,sort&compress]{natbib}
\usepackage[nottoc]{tocbibind}
\usepackage{xcolor}

\newcommand{\pl}{\partial}
\newcommand{\plb}{\bar{\partial}}
\newcommand{\be}{\begin{equation}}
\newcommand{\ee}{\end{equation}}
\newcommand{\bea}{\setlength\arraycolsep{2pt} \begin{eqnarray}}
\newcommand{\eea}{\end{eqnarray}}

\newcommand{\fud}[2]{{}^{#1}{}_{#2}\,}
\newcommand{\fdu}[2]{{}_{#1}{}^{#2}\,}
\newcommand{\fudu}[3]{{}^{#1}{}_{#2}{}^{#3}\,}

 \newcommand{\bry}{{{\bar{y}}}}
 
 \newcommand{\Sigmab}{\bar{\Sigma}}

\newcommand{\strA}[1]{[1]_{#1}}
\newcommand{\strB}[1]{[2]_{#1;}{}}
\newcommand{\strC}[1]{[3]_{#1;}{}}
\newcommand{\strCxi}[1]{{[3]^{\xi}_{#1;}}{}}
\newcommand{\strD}[1]{[4]_{#1;}{}}
\newcommand{\strDxi}[1]{{[4]^{\xi}_{#1;}}{}}

\newcommand{\acoef}{{\mathsf{a}}}

\newcommand{\besubeqs}{\begin{subequations}}
\newcommand{\esubeqs}{\end{subequations}}

\newcommand{\Tr}{\mathrm{Tr}}

\begin{document}
\pagenumbering{gobble}
\hfill
\vskip 0.01\textheight
\begin{center}
{\Large\bfseries 
Higher-spins on Taub-NUT and higher-spin Taub-NUT }

\vskip 0.03\textheight
\renewcommand{\thefootnote}{\fnsymbol{footnote}}
Evgeny \textsc{Skvortsov}\footnote{Research Associate of the Fund for Scientific Research -- FNRS, Belgium}\footnote{Also at Lebedev Institute of Physics}\footnote{\label{note1}First authors, in alphabetical order.}${}^{a}$ \& Yihao \textsc{Yin}\footref{note1}${}^{b,c}$
\renewcommand{\thefootnote}{\arabic{footnote}}
\vskip 0.03\textheight

{\em ${}^{a}$ Service de Physique de l'Univers, Champs et Gravitation, \\ Universit\'e de Mons, 20 place du Parc, 7000 Mons, 
Belgium}\\ \vspace*{5pt}
{\em ${}^{b}$ College of Physics, Nanjing University of Aeronautics and Astronautics, \\
Nanjing 211106, China}\\
{\em ${}^{c}$ Key Laboratory of Aerospace Information Materials and Physics (NUAA), MIIT,\\
Nanjing 211106, China}\\

\vskip 0.05\textheight

\begin{abstract}

We consider higher-spin extensions of self-dual Yang-Mills (HS-SDYM) and self-dual Gravity (HS-SDGR), which are also truncations of chiral higher-spin gravity. Higher-spin fields can consistently propagate on gravitational instantons and we construct solutions to the higher-spin equations on the Taub-NUT background. We show that these solutions remain exact solutions of HS-SDYM. For HS-SDGR the perturbation theory converges for any given spin to give a higher-spin generalization of Taub-NUT and we identify a commutative non-associative algebra that governs the structure of the perturbation theory. 
\end{abstract}

\end{center}
\newpage
\tableofcontents
\newpage
\section*{Introduction}
\setcounter{footnote}{0} 
\pagenumbering{arabic}
\setcounter{page}{1}

At least at present, massless higher-spin states do not show any preference for any value of the cosmological constant. For example, there is a one-to-one correspondence between cubic interactions/amplitudes in Minkowski and (anti)-de Sitter spaces \cite{Bengtsson:1986kh,Metsaev:2018xip}. All well-defined higher-spin gravities, i.e. the numerous $3d$ ones with massless and conformal fields \cite{Blencowe:1988gj,Bergshoeff:1989ns,Pope:1989vj,Fradkin:1989xt,Grigoriev:2019xmp,Grigoriev:2020lzu}, the $4d$ conformal \cite{Segal:2002gd,Tseytlin:2002gz,Bekaert:2010ky} and the $4d$ chiral \cite{Metsaev:1991mt,Metsaev:1991nb, Ponomarev:2016lrm,Skvortsov:2018jea,Skvortsov:2020wtf,Sharapov:2022awp} can tolerate any sign and value of the cosmological constant, see e.g. \cite{Bekaert:2022poo} for a review. Therefore, it is tempting to use the flat space background as a simple playground for more challenging holographic/cosmological applications. From the S-matrix point of view, massless higher-spin fields in flat space should display trivial scattering, see e.g. \cite{Weinberg:1964ew}, but this, as in self-dual theories, results from a nontrivial cancellation of diagrams \cite{Skvortsov:2018jea} (see, however, \cite{Tran:2022amg,Adamo:2022lah} for some exceptions to this lore).

There are many ways to introduce Lorentz-covariant fields that carry the degrees of freedom of massless higher-spin particles. However, many of them are obstructed early on. For example, within the most natural description based on totally symmetric tensor fields $\Phi_{\mu_1...\mu_s}$, called Fronsdal fields \cite{Fronsdal:1978rb}, it is impossible to construct two-derivative gravitational interactions \cite{Aragone:1979hx} and put the fields on any background but the maximally symmetric spaces. At the same time, such cubic gravitational interactions do exist \cite{Bengtsson:1986kh,Metsaev:2018xip}. The chiral formulation \cite{Krasnov:2021nsq}, which originates from twistor theory and where positive and negative helicities are treated differently, is more flexible and allows one to put massless higher-spin fields on any self-dual background, which leads to self-duality playing an important role for massless higher-spin fields. 

It turns out that both self-dual Yang-Mills (SDYM) and self-dual gravity (SDGR) have simple higher-spin extensions  \cite{Ponomarev:2017nrr} where fields interact either via one-derivative interactions of Yang-Mills type (HS-SDYM) or via two-derivative interactions of gravitational type (HS-SDGR). These higher-spin self-dual theories, HS-SDYM and HS-SDGR, admit simple formulations in the light-cone gauge \cite{Ponomarev:2017nrr,Monteiro:2022xwq,Neiman:2024vit} or in terms of chiral fields \cite{Krasnov:2021nsq}, which are straightforward generalizations of SDYM and SDGR, see also \cite{Tran:2021ukl,Tran:2022tft,Adamo:2022lah,Neiman:2024vit,Tran:2025uad,Mason:2025pbz}. In addition, HS-SDYM and HS-SDGR are truncations of chiral higher-spin gravity. All of these theories are smooth in the cosmological constant and within the AdS/CFT context chiral higher-spin gravity is dual to a closed subsector of Chern-Simons vector models \cite{Sharapov:2022awp,Jain:2024bza,Aharony:2024nqs}. 

(Higher-spin) self-dual theories are useful toy models for a great number of reasons out of which we mention a few: they are integrable theories \cite{Ponomarev:2017nrr} with simple formulations, especially on twistor space (see \cite{Tran:2021ukl,Herfray:2022prf,Adamo:2022lah,Mason:2025pbz} for HS-SDYM and HS-SDGR); all solutions are also solutions of the full theories as well as amplitudes are. Lastly, the full theories can be understood as perturbations over the self-dual subsector.

Complete finite-action solutions of SDYM and SDGR are usually called instantons. Likewise, complete finite-action solutions of chiral higher-spin gravity and of HS-SDYM, HS-SDGR can be called higher-spin instantons. In a yet not understood way higher-spin instantons should combine the usual Yang-Mills and gravitational instantons and, perhaps, enrich them with some genuine higher-spin moduli. In \cite{Skvortsov:2024rng}, it was shown that the simplest BPST-instanton \cite{Belavin:1975fg} can be uplifted to chiral theory where it switches on some other low-spin fields. However, the BPST-instanton (one-instanton solution) is too symmetric to trigger any higher-spin fields. Recently, a class of pp-wave solutions of chiral theory was constructed in \cite{Tran:2025yzd}. 

In this paper, in an attempt to chart out the space of higher-spin instantons more, we consider a higher-spin extension of Taub-NUT black-holes \cite{Taub:1950ez,Newman:1963yy} within HS-SDYM and HS-SDGR, which are ``toy models'' for chiral higher-spin gravity. Apart from trying to understand the space of solutions of a theory, an additional, a purely higher-spin argument, is that black-hole-type solutions play an important and surprising role in holography since they are dual to a generating function of the single-trace operators, see \cite{David:2020fea,Lysov:2022nsv,Lysov:2022zlw}. Another idea, see e.g. \cite{Iazeolla:2011cb}, is that higher-spin fields/higher-derivative interactions can smoothen black-hole singularities. 

In Section \ref{sec:gravity}, we begin by recalling the Gibbons-Hawking ansatz and gravitational instantons. Apart from fixing some notation we recast the ansatz in terms of variables suitable for the self-dual gravity, see e.g. \cite{Herfray:2016qvg,Krasnov:2016emc,Krasnov:2021cva}. Next, in Section \ref{sec:freeHS} we consider free higher-spin fields on gravitational instanton backgrounds, a problem already impossible within the Fronsdal formulation, where we give explicit solutions for Taub-NUT. The main results are in Section \ref{sec:HSBH}, where we first review the simple actions for HS-SDYM and HS-SDGR and then proceed to constructing higher-spin extensions of Taub-NUT. 

In the case of HS-SDYM the linearized solutions turn out to be exact, which is not rare, in general.\footnote{For example, the Kerr black hole solution in the Kerr-Schild form solves both free and full nonlinear Einstein equations. Also, the recently constructed family of pp-wave solutions has this property \cite{Tran:2025yzd}.} It is also easy to explain this effect by noticing that the gauge interactions of HS-SDYM do not see the gravitational ``nature'' of Taub-NUT. On the contrary, the gravitational interactions\footnote{By gravitational/gauge interactions we mean those that are two/one-derivative in the light-cone gauge. They are natural higher-spin generalization of the usual ones. This is also confirmed by the fact that it is with these two types of interactions one can have consistent theories, HS-SDGR/HS-SDYM. Turning on a genuine higher-spin interaction would lead to chiral theory as a completion. } of HS-SDGR are binding enough to trigger backreaction. However, two fields with helicities $s_{1,2}>0$ source a field with helicity $s_1+s_2-2$. Therefore, the perturbation theory converges for any fixed spin. The most general solution is parameterized by the numerical factors in front of the free parts. We identify a commutative non-associative algebra $\mathcal{A}$ that governs the perturbation theory. For negative helicity fields the situation is the opposite as they, in principle, can receive infinitely many contributions from fields with even higher spins. They form a representation of $\mathcal{A}$. We conclude with some remarks in Section \ref{sec:conclusions}, after which a number of appendices provide further details of the calculations.

\section{Gravitational instantons: Taub-NUT, ...}
\label{sec:gravity}
Let us introduce the $4d$ coordinates $x^\mu$, which we convert into the spinorial language as $x^{AA'}$.\footnote{The indices are raised and lowered following the Penrose-Rindler convention, i.e. $v^A=\epsilon^{AB}v_B$, $v_B=v^A\epsilon_{AB}$, \textit{idem}. for the primed indices and $\epsilon_{AB}=-\epsilon_{BA}$ is the invariant tensor. If some spin-tensor expression is symmetric or is to be symmetrized in certain indices, we often denote all such indices by the same letter, e.g. $\xi^{AA}$.} In addition, we split $x^{AA'}=t \epsilon^{AA'} + \xi^{AA'}$, where $\xi^{AB}=\xi^{BA}$ are $3d$ coordinates.\footnote{Here we have used the same $2d$ Levi-Civita $\epsilon^{AA'}$ as $\epsilon^{AB}$, $\epsilon^{A'B'}$. Since the ansatz breaks the $4d$ Lorentz symmetry down to the $3d$ one, there has to be an additional vector $v^{AA'}$ to define the $3+1$ split. We have chosen $v^{AA'}=\epsilon^{AA'}$ as matrices, which simplifies the conversion between primed and unprimed indices. Indeed, one would have to define $\bar\xi^{A'}=\xi^B v\fdu{B}{A'}$ for generic $v^{AA'}$, but $\bar\xi^{A'}=\xi^B \epsilon\fdu{B}{A'}=\xi^{A'}$ with our choice and one can avoid introducing $\bar\xi$ notation.} A convenient starting point is the Gibbons-Hawking ansatz for the vierbein \cite{Gibbons:1978tef}
\begin{align}
    e^{AA'}&= \frac{1}{\sqrt{V}}\epsilon^{AA'} (dt + h_{AB}\,d\xi^{AB}) + \sqrt{V}\, d\xi^{AA'}\,,
\end{align}
which gives the metric
\begin{align}
    ds^2&= V^{-1} (dt+h\cdot dx)^2+ V d\xi^2\,.
\end{align}
Here, $V=V(\xi)$ and $h_{AB}=h_{AB}(\xi)$ is the vector potential in $3d$ that depends on $\xi^{AB}$ only. The next step is to solve for the spin-connection. The honest torsion constraint is
\begin{align}
     de^{AA'}&= \varpi\fud{A}{C}\wedge e^{CA'}+\varpi\fud{A'}{C'}\wedge e^{AC'}\,.
\end{align}
For self-dual solutions $\varpi^{AB}$ is pure gauge and can be set to zero, which we immediately do. It should not be confused with another field with a similar name $\omega^{AB}$ below. Sometimes it is useful to perform the $3+1$ decomposition of various differential forms and to this end we can split the vierbein as
\begin{align}
    e^{AA'}&=\epsilon^{AA'} e_0+ e_3^{AA'}\,, & e_0&= \frac{1}{\sqrt{V}} (dt + h_{AB}\,d\xi^{AB})\,, && e_3^{CC}=\sqrt{V}\, d\xi^{CC}\,.
\end{align}
The spin-connection is found to have the following form
\begin{align}
    \varpi^{A'A'}&= g^{A'A'} e_0 + g\fud{A'}{C} e_3^{A'C} =g\fud{A'}{C} e^{A'C}\,,
\end{align}
where
\begin{align}
    g_{CC}&= -V^{-3/2} \partial_{CC} V=-2 V^{-3/2} \pl_{CM}h\fud{M}{C}\,.
\end{align}
The latter equality implies the Bogomolny equation\footnote{One may wonder if one can just take any Yang-Mills instanton, e.g. the BPST solution, and map it to a gravitational one. No such map seems to exit in general, but it is possible to do so if there are two commuting killing vector fields \cite{Tod:2024zpa}. However, one can go the opposite direction: any gravitational instanton gives a Yang-Mills one for the same metric. We are grateful to Kirill Krasnov for the comment.}
\begin{align}
    \partial_{CC} V=2\pl_{CM}h\fud{M}{C}\,,
\end{align}
i.e. $dV=*dh$ or $\pl_i V= \epsilon\fdu{i}{jk}\pl_j h_k$, $i,j,k=1,...,3$. Out of the vierbein one can construct anti-self-dual and self-dual two-forms
\begin{align}
    \Sigma^{AA}&= e\fud{A}{C'}\wedge e^{AC'}= +2e_3^{AA}\wedge e_0 +E^{AA}\,,\\
    \bar\Sigma^{A'A'}&= e\fdu{C}{A'}\wedge e^{CA'}= -2e_3^{A'A'}\wedge e_0 +E^{A'A'}\,,
\end{align}
where $E^{AA}=e_3{}\fud{A}{C}\wedge e_3^{AC}$. It is easy to see that $\Sigma^{AB}=e\fud{A}{C'}\wedge e^{BC'}$ are closed anti-self-dual two-forms, $d\Sigma^{AB}=0$. They obey $\Sigma^{AA}\wedge \Sigma^{AA}=0$ and, hence, define a hyper-K\"{a}hler structure. It is also possible to represent $\Sigma^{AA}$ in a locally exact form:
\begin{align}
    \Sigma^{AA}&=d\omega^{AA}\,,
\end{align}
but we postpone solving this equation to Section \ref{subsec:SDYMSDGR}.

It is time to have a look at the (self-dual component of the) Weyl tensor $\Psi^{A'B'C'D'}$, which determines the Petrov type of the solution. It is defined via
\begin{align}\label{antiweyl}
    d\varpi^{A'A'}&= \varpi\fud{A'}{C'}\wedge \varpi^{A'C'} +\bar\Sigma_{C'C'} \Psi^{A'A'C'C'}\,.
\end{align}
The latter equation also imposes the Einstein equation, which requires $\square V=0$, where $\square \equiv \tfrac12 \pl^\xi_{AB}\pl_\xi^{AB}$. 

To summarize, gravitational instantons solve Einstein equations and have vanishing anti-self-dual component of the Weyl tensor, $\Psi^{AAAA}=0$. The self-dual Weyl tensor, as can be derived from \eqref{antiweyl}, is
\begin{align}\label{weyl}
    \Psi^{A'A'A'A'}&= -\tfrac32 V^{-3} \pl_{A'A'}V\pl_{A'A'}V+\tfrac12 V^{-2} \pl_{A'A'}\pl_{A'A'}V\,.
\end{align}
It satisfies the Bianchi identities
\begin{align}
    \nabla_{MM'}\Psi^{A'A'A'M'}=0\,.
\end{align}
The latter equation has a straightforward higher-spin generalization \cite{Penrose:1965am}, which we discuss later. Assuming that there is no $t$-dependence, the covariant derivative acts as\footnote{One trick here is that $\pl^\xi_{CC'}$, being symmetric, picks the $e_3^{CC'}$ component of $e^{CC'}$. }
\begin{align}
    e^{CC'}\nabla_{CC'}\chi^{B'}&= e^{CC'}\left(V^{-1/2}\pl_{CC'}^\xi \epsilon\fdu{B'}{A'}-\tfrac12(g\fdu{C'}{A'} \epsilon_{CB'} + \epsilon\fdu{C}{A'} g_{B'C'})\right)\chi^{B'}
\end{align}
for any $\chi^{B'}$. For an unprimed object $\chi^B$ that does not depend on $t$ the covariant derivative is just
\begin{align}
    e^{CC'}\nabla_{CC'}\chi^{B}&= e^{CC'}\left(V^{-1/2}\pl_{CC'}^\xi \epsilon\fdu{B}{A}\right)\chi^{B}\,.
\end{align}
The main advantage of the $t$-independent solutions is that we do not have to know $h_{AB}\,d\xi^{AB}$ as it remains hidden in the vierbein.

\paragraph{Petrov type and Instantons.} In order to have type-D the Weyl tensor should factorize as
\begin{align}
    \Psi^{A'A'A'A'}&= \psi_{A'A'}\psi_{A'A'}\,,
\end{align}
where $\psi_{A'A'}$ is some symmetric bi-spinor. The latter can always be factorized as $\psi_{A'A'}=\xi_{A'}\eta_{A'}$. The first term of $\Psi$ \eqref{weyl} is already type-D, but the second one is so only if 
\begin{align}
    \pl_{AA}\pl_{AA}V&= v_{AA} v_{AA} V
\end{align}
for some $v_{AA}$ (of course, the factors of the sum is not the product of the factors and, hence, we also assume that $\pl_{AA}V\sim v_{AA}$ to have the same factors). Taub-NUT black-hole has type-D and the usual choice for harmonic $V$ is
\begin{align}
    V&= v_0+ \frac{m}{|\xi|}=v_0 + \frac{m}{\sqrt{\rho}}\,,
\end{align}
where $\rho\equiv\xi^2\equiv \tfrac12 \xi_{AA}\xi^{AA}$ or we can replace $\xi$ by $\xi-a$ everywhere, where $a$ is the position of the center. Other simple solutions are the multi-center instantons
\begin{align}
    V&= v_0+ \sum_{i=1}^{N}\frac{m_i}{|\xi-a_i|}\,,
\end{align}
which are not type-D unless $N=1$. Another special solution is the Eguchi-Hanson instanton \cite{Eguchi:1978gw,Eguchi:1978xp}, for which $v_0=0$ and $N=2$.

\section{Free higher spin fields on gravitational instantons}
\label{sec:freeHS}
In this section we study free higher-spin fields on the background of gravitational instantons, specifically on Taub-NUT.

\paragraph{Free higher-spin fields.} A massless field of, say, positive helicity\footnote{The formulas below work in any signature and it is usually wise to extend them to a complexified spacetime until some reality conditions are really needed. The notion of helicity appeals to the Minkowski signature, while instantons are solutions of Euclidean theories. Nevertheless, we will refer to fields by their helicity.} $+s$ can be described \cite{Penrose:1965am} by a symmetric rank-$2s$ spin-tensor $\Psi^{A'(2s)}$ that satisfies
\begin{align}\label{penr}
    \nabla\fud{M}{M'}\Psi^{A'(2s-1)M'}=0\,.
\end{align}
We already know one solution for $s=2$, \eqref{weyl}. The negative helicity $-s$ field can be described by a ``conjugate'' spin-tensor $\Psi^{A(2s)}$ that obeys
\begin{align}\label{Psieq}
    \nabla\fdu{M}{B'} \Psi^{A(2s-1)M}&=0\,.
\end{align}
This suffices for free fields, but interesting interactions, e.g. gravitational and Yang-Mills interactions, require gauge potentials. At this moment the description becomes asymmetric with respect to the helicities since it is sufficient to introduce the gauge potential for, say, positive helicities only \cite{Hitchin:1980hp,Woodhouse:1985id}. The gauge potential is $\Phi_{A(2s-1),A'}$ and satisfies 
\begin{align}\label{Phieq}
    \nabla\fdu{A}{M'}\Phi_{A(2s-1),M'}&=0
\end{align}
that is gauge-invariant under
\begin{align}\label{deltaPhi}
    \delta\Phi_{A(2s-1),A'}&= \nabla_{AA'} \zeta_{A(2s-2)}
\end{align}
provided $\nabla\fdu{A}{M'}\nabla_{AM'} \chi_A=0$ for any spinor $\chi_A$. The latter implies that the background $\nabla$ is self-dual. Indeed, $\nabla\fdu{A}{M'}\nabla_{AM'}$ selects a part of the commutator $[\nabla_{AA'},\nabla_{BB'}]$ that contains the anti-self-dual component of the Weyl tensor $\Psi^{ABCD}$, $\nabla\fdu{A}{M'}\nabla_{AM'} \chi_A\sim \Psi_{AAAB}\chi^B$ (the traceless Ricci tensor and the scalar curvature disappear from this particular projection). Therefore, for the equations to be gauge-invariant we need $\Psi^{ABCD}=0$. 

The gauge potential $\Phi_{A(2s-1),A'}$ is also related in a simple way to the Penrose equation \eqref{penr}: $\Psi^{A'(2s)}$ can be interpreted as a gauge-invariant field-strength
\begin{align}
    \Psi^{A'(2s)}&= \overbrace{ \nabla\fdu{A}{A'}...\nabla\fdu{A}{A'}}^{2s-1} \Phi^{A(2s-1),A'}\,.
\end{align}
We will not use the field-strength, since the important interactions, gauge and gravitational ones, we are going to consider, require the gauge potential.

The gauge potential can be replaced by a one-form connection $\omega^{A(2s-2)}$ \cite{Hitchin:1980hp,Krasnov:2021nsq} provided the action/equations of motion are invariant under
\begin{align}\label{gaugesymmetry}
     \delta \omega^{A(2s-2)}&= \nabla \zeta^{A(2s-2)} +e\fud{A}{C'} \eta^{A(2s-3),C'}\,,
\end{align}
where in addition to $\zeta^{A(2s-2)}$ there is a new gauge parameter $\eta^{A(2s-3),A'}$ to eliminate the extra components of $\omega^{A(2s-2)}$ as compared to $\Phi_{A(2s-1),A'}$. Here, $\nabla = e^{CC'}\nabla_{CC'}$ is the one-form Lorentz-covariant derivative. Indeed, one can decompose $\omega^{A(2s-2)}$ as
\begin{align}\label{PhiintoOmega}
    \omega^{A(2s-2)}&= e_{CC'}\Phi^{A(2s-2)C, C'}+ e\fud{A}{C'}\Upsilon^{A(2s-3),C'}
\end{align}
and the job of $\eta^{A(2s-3),C'}$ is to make $\Upsilon$ non-propagating.  There is a simple gauge-invariant action \cite{Krasnov:2021nsq}
\begin{align}\label{niceaction}
    S= \int \Psi^{A(2s)}\wedge \Sigma_{AA}\wedge \nabla \omega_{A(2s-2)}\,, &&\Sigma_{AA}\equiv e_{AB'}\wedge e\fdu{A}{B'}\,,
\end{align}
which is gauge invariant on any self-dual background, i.e. $\nabla^2 \chi^A\sim \Sigma\fud{A}{B}\chi^B$, see the discussion after \eqref{deltaPhi}.\footnote{In more detail: $\nabla^2 \chi^A\sim \Sigma\fud{A}{B}\chi^B +\bar{\Sigma}_{C'C'}R\fudu{A}{B}{,C'C'}\chi^B+\Sigma_{CC}\Psi\fud{CCA}{B}\chi^B$. The first (cosmological) term will disappear in the action due to one more $\Sigma_{AA}$ in there, as well as the traceless Ricci tensor, $\Sigma^{AB}\wedge \bar{\Sigma}^{A'B'}\equiv0$. } Therefore, one can consistently put massless higher-spin fields on any gravitational instanton background. The equation for $\omega^{A(2s-2)}$ is
\begin{align}
    \Sigma^{AA}\wedge \nabla\omega^{A(2s-2)}&=0\,,
\end{align}
and is equivalent to \eqref{Phieq}. It is also equivalent to (thanks to $\Sigma^{AA}\wedge e^{AC'}\equiv0$)
\begin{align}
    d\omega^{A(n)}&=e\fud{A}{B'}\wedge \omega^{A(n-1),B'}\,,
\end{align}
where we have introduced another one-form field $\omega^{A(n-1),B'}$. The equation for $\Psi$ is
\begin{align}\label{Psieqagain}
    \Sigma_{BB} \wedge d\Psi^{A(2s-2)BB}&=0\,,
\end{align}
and is equivalent to \eqref{Psieq}. Let us find solutions for the equations above on the Taub-NUT background.

\paragraph{Taub-NUT and not only, Negative helicity.} The simplest case is that of the negative helicity fields, where, assuming $\Psi=\Psi(\xi)$, we need to solve
\begin{align}\label{penrconjugate}
    \nabla\fdu{M}{B'} \Psi^{A(2s-1)M}=V^{-1/2}\pl^\xi\fdu{M}{B'} \Psi^{A(2s-1)M}&=0 
\end{align}
i.e. the solutions are as if in the $3d$ flat space. This applies to any Gibbons-Hawking metric, not necessarily Taub-NUT. It is now convenient to package all spins into a single generating function ($y^A$ are auxiliary commuting spinors)
\begin{align}
    \Psi(y)&=\sum_s \Psi_{A(2s-1)} \, y^A...y^A\,. 
\end{align} 
As an ansatz for $\Psi$ we have nothing to think of but 
\begin{align}\label{psiansatz}
    \Psi^{A(2s)}&= \kappa_s(\rho) \xi^{AA}...\,\xi^{AA} && \longleftrightarrow && \Psi= \kappa_s u^s\,,
\end{align}
where $u\equiv y^A y^B \xi_{AB}\equiv (y\xi y)$ will be used many times in what follows. To find $\kappa_s$ we need 
\begin{align}
    d\Psi&= e^{AA'}\nabla_{AA'} \Psi= V^{-1/2} e^{AA'}(s \kappa_s u^{s-1} y_A y_{A'} + \kappa_s' \xi_{AA'} u^{s})\,.
\end{align}
This needs to be projected onto the right component, which is achieved with the help of $\pl_B\nabla\fud{B}{A'}\Psi$, where $\pl_A\equiv \pl/\pl y^A$. The solution to the free equations can be obtained from
\begin{align}\label{solvingPsi}
    \pl_B\nabla\fud{B}{A'}\Psi&= V^{-1/2} u^{s-1}y_{A'}s ((2s+1) \kappa_s+ 2 \kappa_s'\rho)=0\,.
\end{align}
We find $\kappa_s\sim \rho^{-s-1/2}$.

\paragraph{Taub-NUT, Positive helicity.} For the positive helicity fields the spin-connection enters the game and it does not seem easy to solve the equations for arbitrary Gibbons-Hawking metrics. Again, we choose to package all fields into a generating function ($\bry^{A'}$ is one more auxiliary spinor)
\begin{align}
    \Phi(y,\bry)&=\sum_s \Phi_{A(2s-1),A'} \, y^A...y^A\, \bry^{A'} \,.
\end{align}
Likewise, a generating function for $\omega^{A(2s-2)}$ can be defined as\footnote{This gives a slightly different normalization as compared to \eqref{PhiintoOmega}.}
\begin{align}\label{omegaPhi}
    \omega&=e^{CC'}\pl_C\plb_{C'}\Phi\,.
\end{align}
A simple solution can be found by using the data we already have in the Taub-NUT background, i.e. let us take the ansatz 
\begin{align}
    \Phi(y,\bry)&= u^{s-1} (y\xi \bry) \sigma_1(\rho) + u^{s-1} w \sigma(\rho)\,,
\end{align}
where $u\equiv (y\xi y)$ and $w\equiv (\bry y)\equiv \bry^C y_C$.\footnote{According to our convention $\bar y^{C}= \bar y^{C'} \epsilon\fdu{C'}{C}$, which allows us to define $\bry^C y_C$ as if we just replaced $C'$ with $C$. From the $3d$ point of view, both $\bry^{C'}$ and $y^C$ transform in the same representation.} The first term is pure gauge. Indeed, for the gauge parameter \eqref{gaugesymmetry}, which is independent of $\bry$, we can write
\begin{align}
    \zeta(y)= \sigma_2(\rho) u^{s-1}\,.
\end{align}
The second term with the indices released is $\sigma \xi^{AA}...\xi^{AA}\epsilon^{AA'}$. The covariant derivative acts as $\nabla_{CC'}=V^{-1/2}\pl^\xi_{CC'}$ and the first term of $\Phi$ can be eliminated with the help of $\zeta$ above. To get an equation for $\sigma$ we note that
\besubeqs
\begin{align}
    \nabla_{CC'}\sigma(\rho)&= V^{-1/2} \sigma' \xi_{CC'}\,, \\
    \nabla_{CC'} u&=V^{-1/2} y_{C} y_{C'}\,, \\
    \nabla_{CC'}w&= -\tfrac12 g\fdu{C'}{B'}(\bry_{B'} y_C +\bry_{C} y_{B'})\,.
\end{align}
\esubeqs
The projection of $\nabla_{AA'}\Phi$ onto the equation of motion \eqref{Phieq} can be implemented as $y^C\plb_{B'}\nabla\fdu{C}{B'}$, which gives
\begin{align}
   y^C\plb_{B'}\nabla\fdu{C}{B'}\Phi&=u^s[-V^{-1/2} \sigma' +\tfrac12 g \sigma] \,.
\end{align}
Note that the equation does not actually depend on spin, because the spin-connection acts on the single primed index. Recall that $g=-V^{-3/2}V'$. It is easy to solve for $\sigma$ for ``arbitrary'' $V$.\footnote{It is not arbitrary since we have already used $\pl_{AA'}V=V'\xi_{AA'}$ to get the spin-connection.}  We find
\begin{align}\label{linearizedFluc}
    \sigma&= V^{-1/2}
\end{align}
for all spins. This solves for the positive helicity fields on the Taub-NUT background.

\section{Higher spin black holes}
\label{sec:HSBH}
In this Section we first define SDYM and SDGR and the higher-spin extensions thereof and then try to uplift free higher-spin fields on the Taub-NUT instanton to become higher-spin Taub-NUT instantons. 

\subsection{SDYM and SDGR}
\label{subsec:SDYMSDGR}
\paragraph{Self-dual Yang-Mills theory.} The Lorentz covariant action, the Chalmers-Siegel action \cite{Chalmers:1996rq}, for SDYM can be found as follows. One begins with the Yang-Mills action (in the two-component spinorial language)
\begin{align}
    S_{YM}&=\tfrac14 \int F_{\mu\nu}^2 = \tfrac18\int \big(F_{AB}F^{AB}+ F_{A'B'}F^{A'B'}\big)\,,
\end{align}
where we defined
\begin{align}
F_{\mu\nu} \equiv F_{AA'BB'} = \tfrac{1}{2}F_{AB}\epsilon_{A'B'} + \tfrac{1}{2}F_{A'B'}\epsilon_{AB} \,,
\end{align}
and $F_{\mu\nu}F^{\mu\nu}= \tfrac{1}{2}\big(F_{AB}F^{AB}+ F_{A'B'}F^{A'B'}\big)$. We add the right amount of the topological invariant
\begin{align}
    \int F\wedge F \sim\int d^4x\, \Big(F_{AB}F^{AB}- F_{A'B'}F^{A'B'}  \Big)  
\end{align}
to reduce the action to
\begin{align}
    S_{YM}&=\tfrac14 \int F_{AB}F^{AB}\,.
\end{align}
Next, one introduces an auxiliary field $\varphi_{AB}$
\begin{align}
S_{\text{YM}}[A,\varphi] = \int  \mathrm{d}^4 x\, \Big(+\tfrac{1}{2}\varphi_{AB}F^{AB} - \tfrac{1}{4}\varphi_{AB}\varphi^{AB}  \Big) \,.
\end{align}
The action is still equivalent to the standard Yang-Mills action, modulo a total derivative. One can drop the second term, the resulting action is SDYM
\begin{align}
S_{\text{SDYM}}[A,\varphi] = \tfrac{1}{2}\int  \mathrm{d}^4 x\, \varphi_{AB}F^{AB}  \,.
\end{align}
By rescaling the fields one can introduce a parameter $\epsilon$ in front of $\varphi_{AB}\varphi^{AB}$ in $S_{\text{YM}}[A,\varphi]$ and treat Yang-Mills theory as a deformation of SDYM. The last step, which is relevant below, is to allow for an arbitrary gravitational background 
\begin{align}
S_{\text{SDYM}}[A,\varphi] = \tfrac{1}{2}\int  \varphi^{AA}\, \Sigma_{AA}\wedge F  \,.
\end{align}
We recall that $\Sigma_{AA}\equiv e_{AB'}\wedge e\fdu{A}{B'}$. In the free field limit, $F=dA$ and the action is the $s=1$ case of \eqref{niceaction}. 

\paragraph{Self-dual gravity.} The self-dual gravity with vanishing cosmological constant has a simple action \cite{Krasnov:2021cva} (another option is \cite{Siegel:1992wd})\footnote{We do not review a path from the Einstein-Hilbert action to this one, see \cite{Krasnov:2021cva}, but it is just a bit longer route of integrating in/out auxiliary fields as compare to the YM/SDYM relation. }
\begin{align}
    S_{\text{SDGR}}[\Psi, \omega]&=\tfrac12\int \Psi^{AAAA}\, d\omega_{AA}\wedge d\omega_{AA} \,.
\end{align}
Here, $\Psi^{AAAA}$ is reminiscent of the anti-self-dual part of the Weyl tensor and $\omega^{AA}$ is a one-form, which is reminiscent of the anti-self-dual component of the spin-connection. 
Let us explain why this action leads to the equations we went through in Section \ref{sec:gravity}. The equation of motion for $\Psi$ implies that  $\Sigma^{AA}\wedge \Sigma^{AA}=0$, where $\Sigma^{AA}=d\omega^{AA}$. This means that $\Sigma^{AA}=e\fud{A}{B'}\wedge e^{AB'}$ for some $e^{AA'}$. $d\Sigma^{AA}=0$ implies the half-torsion constraint
\begin{align}
     de^{AA'}&= \varpi\fud{A'}{C'}\wedge e^{AC'}
\end{align}
for some $\varpi^{A'A'}$. Thus, we have recovered the field content introduced in Section \ref{sec:gravity}. When linearized over some background $\Sigma^{AA}=d\omega^{AA}_0$, the action reduces to 
\begin{align}
    S&=\int \Psi^{AAAA}\, \Sigma_{AA}\wedge d\omega_{AA} \,,
\end{align}
which is the $s=2$ case of the free action \eqref{niceaction}. The equation of motion for the fluctuation $\omega^{AA}$ implies the known equation for $\Psi$, \eqref{Psieq}, in the form $\Sigma_{AA}\wedge d\Psi^{AAAA}=0$, \eqref{Psieqagain} (note that the spin-connection with unprimed indices vanishes). 

\paragraph{Potential for Taub-NUT.} Let us fill in one gap. We already know the vierbein and, hence, $\Sigma^{AA}$ for Taub-NUT. Solving the equation $d\omega^{AA}_0=e\fud{A}{B'}\wedge e^{AB'}=\Sigma^{AA}$ with the Taub-NUT vierbein, we obtain a simple solution (not unique) for $\omega^{AA}$:
\begin{align}\label{potentialTaub}
\omega^{AA}_0&=\phi^{AAA,A'}e_{AA'}+\phi^A{}_{A'}e^{AA'} \ ,
\end{align}
where (with $V=v_0+\frac{m}{\sqrt{\rho}}$)
\begin{align}
\phi^{AAA,A'}=\xi^{AA}\epsilon^{AA'}\frac{v_0}{2\sqrt{V}}
\ \ \ \text{and} \ \ \ 
\phi^{AA'}=\xi^{AA'}\left(2\sqrt{V}-\frac{2v_0}{3\sqrt{V}}\right) \ .
\end{align}

\paragraph{How black hole instantons fluctuate.} Since Taub-NUT is an instanton one should expect that all fluctuations correspond to changes in the free parameters that determine the instanton. Let us recall that with the Gibbons-Hawking ansatz 
\begin{align}
    V&= v_0 +\frac{m}{|\xi-a|}\,.
\end{align}
Starting from the exact solution of $\omega ^{AA}_0$ above, we can derive%
\begin{equation}
\omega ^{AA}_0=2\xi ^{AA}\left( dt+h_{BB}d\xi ^{BB}\right) +2V\xi
^{A}{}_{A^{\prime }}d\xi ^{AA^{\prime }}-v_{0}\xi ^{A}{}_{A^{\prime }}d\xi
^{AA^{\prime }}\ ,
\end{equation}%
then, using $\delta V=\delta v_{0}$, we find
\begin{eqnarray}
\delta \omega ^{AA}_0 &=&\left( \delta v_{0}\right) \xi ^{A}{}_{A^{\prime
}}d\xi ^{AA^{\prime }}  \notag \\
&=&\left( \delta v_{0}\right) V^{-\frac{1}{2}}\left( -\tfrac{1}{2}\xi
^{AA}\epsilon ^{AA^{\prime }}e_{AA^{\prime }}+\tfrac{2}{3}\xi
^{A}{}_{A^{\prime }}e^{AA^{\prime }}\right) \ .\label{flucta}
\end{eqnarray}
Now, we recall that we found in \eqref{linearizedFluc} a class of linearized fluctuations, which for $s=2$ reads
\begin{align}\label{spintwofluc}
    \phi^{AAA,A'}&\sim V^{-1/2} \xi^{AA}\epsilon^{AA'}\,, & \phi^{A,A'}&=0\,.
\end{align}
We observe that this fluctuation corresponds to a variation of $v_0$ plus a certain diffeomorphism. Note that $\phi^{A,A'}$ can be eliminated with the help of the algebraic (shift or Stueckelberg) symmetry, which is the last term in \eqref{gaugesymmetry}. For $s=2$ this corresponds to diffeomorphisms. 
The first term in \eqref{flucta} is exactly the linearized fluctuation \eqref{spintwofluc}, which satisfies the free equation of motion by itself. Therefore, we can assume that the gravitational background is kept fixed.

\subsection{Higher-spin extension of SDYM}
\label{sec:HSSDYM}

\paragraph{HS-SDYM.} There is a simple higher-spin extension of SDYM. 
\begin{align}
    S&=\sum_{m}\Tr\int \Psi^{A(m+2)}\, \Sigma_{AA}\wedge F_{A(m)} \,,
\end{align}
where the curvatures $F_{A(m)}$ are better defined in terms of generating functions as
\begin{align}
    F&= d\omega- \tfrac12 [\omega,\omega]\,,
\end{align}
where $[\bullet,\bullet]$ is the Lie bracket, i.e we assume that $\omega\equiv \omega(y)\equiv \omega^a(y)\,T_a$, where $T_a$ are generators of some Lie algebra. Without loss of generality one can take a matrix realization, i.e. $\omega(y)=\omega(y)\fdu{i}{j}$. Then we can simply write $F=d\omega -\omega\omega$, implying the matrix multiplication. Let us note that the $\omega\omega$-piece is just the commutative limit of the Moyal-Weyl star-product for the gluonic chiral higher-spin gravity \cite{Skvortsov:2020wtf,Monteiro:2022xwq}. In principle, one can alter the relative coupling constants between various higher-spin fields, but it makes more sense and is also easier to keep them as in the chiral higher-spin gravity. Let us note that there are further contractions \cite{Serrani:2025owx} of HS-SDYM that can even have finitely-many fields.

\paragraph{Positive helicity.} With the interactions turned on we have for the equations of motion (matrix product is implied) 
\begin{align}
    \Sigma^{AA}\wedge \left(d\omega^{A(n)}-\sum_{k}\, \omega^{A(n-k)}\wedge \omega^{A(k)}\right) &=0 \,.
\end{align}
With the solution \eqref{linearizedFluc} to the free equations we have
\begin{align}\label{sigmabarpart}
    d\omega&= d\omega(\bar{\Sigma})=\tfrac12\bar{\Sigma}^{C'C'}\left(\chi_1^{(s)} u^{s-2} y_{C'}y_{C'}+\chi_2^{(s)} u^{s-1} \xi_{C'C'} +\chi_3^{(s)} u^{s-2} \xi_{C'B}y^B\xi_{C'B}y^B \right)\,,
\end{align}
where we define
\besubeqs
\begin{align}\label{chiabc}
    \chi_1^{(s)}&=-2V^{-1/2}(s-1)(\rho \sigma' +s \sigma) \,, \\
    \chi_2^{(s)}&=-(V^{-1/2}\sigma' +(s+\tfrac12) g\sigma) \,,\\
    \chi_3^{(s)}&=-g \sigma (s-1)\,.
\end{align}
\esubeqs
On the other hand 
\begin{align}
    \omega\wedge \omega&= \sum_{s_1+s_2=s+1} u^{s-2} \tfrac12[\sigma_{s_1},\sigma_{s_2}](s_1-s_2)(\Sigma^{CC} y_C \xi_{CB}y^B-\bar{\Sigma}^{C'C'} y_{C'} \xi_{C'B}y^B) 
\end{align}
where we put the color into $\sigma$, i.e. $\sigma_{s}=\sigma^a T_a$. We observe that
\begin{align}
    y_Ay_A\Sigma^{AA}\wedge d\omega&\equiv0 \,, &
    y_A y_A \Sigma^{AA}\,\omega\wedge \omega&\equiv 0\,.
\end{align}
Therefore, the solution of the free equations is also a solution of the nonlinear ones! As a result, if the higher-spin interactions are of the Yang-Mills type the higher-spin Taub-NUT black-hole is just a collection of free higher-spin fields on the Taub-NUT background.

\paragraph{Negative helicity.} In the first approximation $\Psi$ are just the free negative helicity fields propagating on the Taub-NUT background. In the next approximation they can scatter off the $\omega$-background via
\begin{align}
    \left(d\Psi^{A(2s)}+ \sum_{m} [\Psi^{A(2s)B(m)},\omega_{B(m)}]\right)\wedge\Sigma_{AA}&=0\,.
\end{align}
A somewhat lengthy calculation in Appendix \ref{app:SDYMNegative} for the second term and \eqref{dPsicomputed} for the first reveal that they simply have different tensor structures. Therefore, it is impossible to match them. Note that with the maximal symmetry assumed here (everything depends on $\xi^{AA}$ only) the ansatz \eqref{psiansatz} for $\Psi$ is unique. 

We observe that the tensor structure of the source is inconsistent with the field $\Psi$ it is a source to. In the higher-spin gravity it has so far been possible to introduce color only via Chan-Paton-type approach \cite{Konstein:1989ij,Metsaev:1991nb,Skvortsov:2020wtf}, with the groups being $U(N)$, $O(N)$ and $USp(N)$. While there can be multiple spin-two fields, the true graviton (there can be only one metric to determine the spacetime geometry) is the one that is a singlet under the gauge group: for $U(N)$ it corresponds to $U(1)$ in $SU(N)\times U(1)$, for $O(N)/USp(N)$ even spins are rank-two symmetric/antisymmetric tensors of the gauge group, i.e. are always reducible and contain the trivial representation. In practice, the true graviton is a multiple of the unit matrix. Now, if we associate the higher-spin generalizations of the Taub-NUT solution with the unit matrix as well, the source vanishes. 

HS-SDYM allows for any gauge group (in some sense, the gauge interactions are not as restrictive and allow for other gaugings, while the genuine higher-spin interactions seem to allow only for the Chan-Paton-type gaugings). In this case, one should require $\Psi$ to live in a subspace that commutes to all $\omega_{s}$. To summarize, the source vanishes (can be made to vanish) and the linearized solution $\Psi$ is still a solution.

\subsection{Higher-spin extension of SDGR}
\label{sec:HSSDGR}

\paragraph{HS-SDGR.} Chiral higher-spin gravity admits a truncation to the two-derivative (gravitational) interactions, which has a simple action
\begin{align}
    S&=\tfrac12\sum_{n,m}a_{n,m}\int \Psi^{A(m+n)}\, d\omega_{A(n)}\wedge d\omega_{A(m)} \,.
\end{align}
The coefficients $a_{n,m}$ are not yet fixed above, but they do have specific values $a_{s_1,s_2}=a_{2,s}=1$, when coming from the chiral theory, see Appendix \ref{app:normalization}. Again, there are various contractions \cite{Serrani:2025owx} of HS-SDGR. Let us expand this action over the gravitational instanton $\Sigma^{AA}=d\omega^{AA}_0$
\begin{align}
    S&=\sum_{n,m}a_{2,m}\int \Psi^{A(m+2)}\, \Sigma_{AA}\wedge d\omega_{A(m)} +\tfrac12\sum_{n,m}a_{n,m}\int \Psi^{A(m+n)}\, d\omega_{A(n)}\wedge d\omega_{A(m)} \,.
\end{align}
It will also be useful to introduce a pairing $\langle f| g\rangle $ for any two generating functions $f(y)$, $g(y)$ as
\begin{align}\label{pairing}
    \langle f| g\rangle&= \exp[\pl^B_{1} \pl^2_{B}]\, f(y_1)g(y_2)\Big|_{y_i=0}=f(-\pl) g(y) \Big|_{y=0}=g(\pl) f(y) \Big|_{y=0}\,.
\end{align}
Now, the action can also be written without any explicit indices
\begin{align}
    S&= \int \langle \Psi | \Sigma_{AA} y^A y^A \, d\omega \rangle+\tfrac12 \langle \Psi |  d\omega\wedge d\omega \rangle
\end{align}
The two equivalent forms above can be used to vary the action.

\subsubsection{Positive helicity}
The linearized equation is the same as before, of course,
\begin{align}
    \Sigma^{AA}\wedge d\omega^{A(n)}&=0 \,,
\end{align}
and it has the same solution \eqref{linearizedFluc}. We can add the interactions between higher-spin fields themselves to find
\begin{align}\label{fullhssdgreq}
    a_{2,s}\Sigma^{AA}\wedge d\omega^{A(n)}+\tfrac12 \sum_{k}a_{n+2-k,k}\, d\omega^{A(n+2-k)}\wedge d\omega^{A(k)} &=0 \,.
\end{align}
With the couplings coming from chiral theory, in the language of generating function we find simply
\begin{align}
    y_A y_A \Sigma^{AA}\wedge d \omega(y) +\tfrac12 d\omega(y)\wedge d \omega(y)=0\,.
\end{align}
 As different from HS-SDYM, the source does not vanish. Indeed, with the help of \eqref{sigmabarpart} one finds
\begin{align}
    d\omega\wedge d\omega&= -\tfrac1{12}\mathrm{vol}\sum_{s_1+s_2=s-2}a_{s_1,s_2} \chi^{s_1,s_2}_- u^s\,,
\end{align}
and, hence, we have to correct the free solution. Here, we defined\footnote{The same definitions \eqref{chiabc} apply, but we are going to replace $\sigma$ with $\sigma_s$ that depends on spin and is not necessarily equal to the linearized solution. }
\begin{align}\notag
    \chi^{s_1,s_2}_-&= \left(\chi_1^{(s_1)}\chi_2^{(s_2)}+\chi_2^{(s_1)}\chi_1^{(s_2)}+2\rho\chi_2^{(s_1)}\chi_2^{(s_2)}+\chi_1^{(s_1)}\chi_3^{(s_2)}+\chi_3^{(s_1)}\chi_1^{(s_2)}+\rho(\chi_2^{(s_1)}\chi_3^{(s_2)}+\chi_3^{(s_1)}\chi_2^{(s_2)})\right)
\end{align}
and
\begin{align}
    \mathrm{vol}&= \Sigma_{AA}\wedge \Sigma^{AA}=-\bar{\Sigma}_{A'A'}\wedge \bar{\Sigma}^{A'A'}\,.
\end{align}
A solution to the nonlinear equations can be found by using the data we already have in the Taub-NUT background. Let us take the same ansatz 
\begin{align}
    \omega&=e^{CC'}\pl_C\plb_{C'}\Phi\,, & \Phi(y,\bry)&= u^{s-1} w \sigma_s(\rho)\,, 
\end{align}
as for the linearized solution \eqref{linearizedFluc}, but keep $\sigma_s$ free for a moment, which leads to
\begin{align}\label{omegasplit}
    d\omega&= d\omega(\bar{\Sigma})+d\omega(\Sigma)\,,
\end{align}
where
\besubeqs
\begin{align}
    d\omega(\Sigma)&= \tfrac12{\Sigma}^{CC}\left(\xi_{CB}y^B\xi_{CB}y^B 2(s-1)u^{s-2}+u^{s-1} \xi_{CC}\right)\chi_0^{(s)}\,,\\
    \chi_0^{(s)}&= V^{-1/2}\sigma_s'-\tfrac12 g \sigma_s\,.
\end{align}
\esubeqs
We get an additional contribution to the source of the form
\begin{align}
    d\omega(\Sigma)\wedge d\omega(\Sigma)&= \tfrac{1}{12}\sum_{s_1+s_2=s+2} a_{s_1,s_2} 2u^s \rho (s+1) \chi_0^{(s_1)}\chi_0^{(s_2)} \,.
\end{align}
The linear part of Eq. \eqref{fullhssdgreq} gives
\begin{align}
    y_A y_A \Sigma^{AA}\wedge d\omega&= \tfrac13 \tfrac12  \chi_0^{(s)} (2s-1)u^s\,.
\end{align}
Finally, Eq. \eqref{fullhssdgreq} boils down to, schematically,  
\begin{align}
    y_A y_A \Sigma^{AA}\wedge d \omega(\Sigma) +\tfrac12 d\omega(\Sigma)\wedge d \omega(\Sigma)+\tfrac12 d\omega(\bar\Sigma)\wedge d \omega(\bar\Sigma)=0\,,
\end{align}
where we displayed the terms that do contribute upon splitting $d\omega$ into (anti)self-dual components \eqref{omegasplit}. Concretely, we find
\begin{align}
    \tfrac{\alpha}{6} (2s-1) a_{2,s}\chi_0^{(s)}u^s + \tfrac12\sum_{s_1+s_2=s+2} \tfrac{1}{12} a_{s_1,s_2} (-\chi_{-}^{s_1,s_2} + \chi_0^{(s_1)}\chi_0^{(s_2)}2(s+1)\rho)u^s=0\,,
\end{align}
where we introduced $\alpha$ in $\alpha y_A y_A \Sigma^{AA}$, which is a formal coupling constant and $\alpha^{-1}$ is supposed to be small.

\paragraph{General perturbation theory.} The equation for a spin-$s$ field has a source built from pairs of fields with spins $s_1$, $s_2$, such that $\hat{s}=\hat{s}_1+\hat{s}_2$, where $s=\hat{s}+2$, i.e. $s=s_1+s_2-2$. There is always a linearized solution to \eqref{fullhssdgreq}. We recall that the spin-two sector is frozen to be the background since there are no nontrivial fluctuations there. The equation has a hierarchy structure: lower spin fields can be sources of higher spin fields but not vice versa. For example, $\hat{1}$, i.e. $s=3$, field is always free.\footnote{As always, one can truncate to even spins only, which would eliminate $s=3$ and make $s=4$ free. } Next, $\hat{2}$ is sourced by two $\hat{1}$s. $\hat{3}$ is sourced by $\hat{2}$ and $\hat{1}$, while $\hat{2}$ consists of the free solution and the one obtained via $\hat{1}+\hat{1}$ source. For $\hat{4}$ we have $\hat{1}+\hat{1}+\hat{1}+\hat{1}$, $\hat{2}+\hat{1}+\hat{1}$, $\hat{3}+\hat{1}$, $\hat{2}+\hat{2}$, where again $\hat{2}$ and $\hat{3}$ can also have some linearized pieces. An important consequence of this ``hierarchy'' is that the perturbation theory converges for any given spin.   

To systematize the perturbation theory, let the free solution for a spin-$s$ field have weight $\hat{s}$. The source for a spin-$s$ field can be enumerated by all binary trees with $k\leq \hat{s}$ leaves. The leaves correspond to the initial (free) fields that build the source (by passing through the equations for lower spins). Now, one needs to list all partitions of $\hat{s}$ into exactly $k$ numbers and distribute them over the leaves. The number, say $\hat{n}$, on a leaf corresponds to the free solution for spin $\hat{n}$ equation. There is an ambiguity in how to normalize the free solutions, which leads to infinitely many parameters we denote $c_{s,0}$ for the full nonlinear solutions. Any chosen normalization is accompanied by a small 'coupling' constant $g\sim \alpha^{-1}$ to make sure that the higher-spin fields do not dominate and have a well-defined perturbation theory over the gravitational background.  

The perturbation theory for HS-SDGR has an additional feature that is absent in the chiral theory: for any given spin the expansion converges in the sense that it stops at a certain order, which is thanks to the ``hierarchy'' mentioned above. 

\paragraph{One iteration of perturbation theory.} The free equations are solved by \eqref{linearizedFluc} with $\sigma_{s}\sim V^{-1/2}$. Let us take two such free solutions and plug them into the source term. The resulting equation can be solved with respect to the spin-$s$ field for any values of $a_{s_1,s_2}$ to find
\begin{align}\label{spin4from3}
    \sigma_s&= a_{s_1,s_2}\frac{ s_1 s_2 \left(2 \left(s+1\right) V-v_0\right)}{\alpha \left(2s -1\right) V^{5/2}}\,,
\end{align}
where $s=s_1+s_2-2$. One can notice that this, next-order, solution just contains $V^{-1}$, $V^{-2}$ on top of the free solution. It can easily be checked that the pattern persists and the deeper we go into the perturbation theory the more negative powers of $V$ are generated and nothing else pops up. Therefore, it makes sense to find out what two generic negative powers $\sigma_{s_1}\ni V^{-1/2} V^{-m_1}$ and $\sigma_{s_2}\ni V^{-1/2} V^{-m_2}$ lead to. The symbol $\ni$ is to emphasize that two monomials do not represent a consistent solution yet. These two monomials produce for $\sigma_s$
\begin{align}
    \sigma_s\ni\tfrac{\left(s_1+s_2+3\right) \left(m_1 \left(s_2+2\right)+\left(s_1+2\right) m_2+2\left(s_1+2\right) \left( s_2+2\right)\right) V^{-m_1-m_2-1}}{2\alpha  \left(m_1+m_2+1\right) \left(2 s_1+2 s_2+3\right)}-\tfrac{\left(s_1+2\right) \left(s_2+2\right) v_0 V^{-m_1-m_2-2}}{2\alpha  \left(2 s_1+2 s_2+3\right)}\,.
\end{align}
Here, we just displayed a contribution for a single term in the sum $\sum_{s_1,s_2}$. The complete contribution will be twice that if $s_1\neq s_2$ due to the sum. See Appendix \ref{app:positivehelicity} for more detail. Clearly, with the help of the formula above one can construct the most general perturbative solution by starting with a superposition of free ones.

\paragraph{Commutative non-associative algebra behind instantons.} To formalize what happens in the perturbation theory one can define a commutative but non-associative product for generators $V_s^m$ (here $s$ is actually $\hat{s}$ as the latter is more convenient since it is additive) that reproduces the operation above (from now on we fix $a_{s_1,s_2}=1$), i.e.
\begin{align}
    V_{s_1}^{m_1}\circ V_{s_2}^{m_2} = f^1_{s_1,s_2|m_1,m_2}V_{s_1+s_2}^{m_1+m_2+1}+f^2_{s_1,s_2|m_1,m_2}V_{s_1+s_2}^{m_1+m_2+2}\,, 
\end{align}
where the structure constants are
\besubeqs
\begin{align}
    f^1_{s_1,s_2|m_1,m_2}&=\frac{\left(s_1+s_2+3\right) \left(m_1 \left(s_2+2\right)+\left(s_1+2\right) m_2+2\left(s_1+2\right) \left(s_2+2\right)\right) }{2\alpha  \left(m_1+m_2+1\right) \left(2 s_1+2 s_2+3\right)}\,,\\
    f^2_{s_1,s_2|m_1,m_2}&= -v_0\frac{\left(s_1+2\right) \left(s_2+2\right) }{2\alpha  \left(2 s_1+2 s_2+3\right)}
\end{align}
\esubeqs
This product encodes the complete information about the perturbation theory. The basis element $V_s^0$ corresponds to the free solutions. We see, for example,
\begin{align}
   V_1^0\circ V_1^0&= \frac{45 V_2^1}{7 \alpha }-\frac{9 V_2^2 v_0}{14 \alpha }\,,
\end{align}
which is the spin-$4$ field generated by the source made of two spin-$3$ fields, cf. \eqref{spin4from3}. Suppose we would like to ignore the free parts of all fields except for $s=3$, i.e. $\hat{s}=1$. Then, the generating function of the solution is 
\begin{align}
    V_1^0+V_1^0\circ V_1^0+\text{all possible bracketings of } V_1^0\,.
\end{align}
A general solution is parameterized by the free coefficients in front of the free solutions
\begin{align}
    F&= \sum c_{s,0} V_s^0 \,.
\end{align}
The solution to any order in perturbation theory can be obtained by evaluating the sum of all products of $F$
\begin{align}
    &F+F\circ F + (F\circ (F\circ F))+((F\circ F)\circ F) + (F\circ F)\circ (F\circ F)+...=\\
    &F+(F\circ F)+2((F\circ F)\circ F)+(F\circ F)\circ (F\circ F)+4 ((((F\circ F)\circ F)\circ F))+...
\end{align}
We can also package all orders of perturbation theory into a generating function $F(t)$ that obeys (some combinatorial aspects can be found in \cite{MALHAM2024134054} and refs therein)
\begin{align}
    \pl_t F&=F\circ F\,.
\end{align}

\subsubsection{Negative helicity}
The linearized equation for $\Psi$ reads as before
\begin{align}
    \Sigma_{AA}\wedge d\Psi^{A(n)}&=0 
\end{align}
and was solved in \eqref{solvingPsi}, which we recall for the spin-$s$ field to be
\begin{align}
    \Psi&= \kappa_s(\rho) u^s\,, && \kappa_s\sim \rho^{-1/2-s}\,.
\end{align}
As for the positive helicity case, in order to account for interactions, we allow $\kappa_s$ to deviate from the solution above. The interactions lead to
\begin{align}
    a_{2,m}d\Psi^{A(m+2)}\wedge \Sigma_{AA}+\sum_{n}a_{n,m}\,d\Psi^{A(m+n)}\wedge d\omega_{A(n)}&=0\,.
\end{align}
It is more efficient to use the definition of the pairing \eqref{pairing} to present the equations as (assuming the chiral theory's normalization)
\begin{align}
    \left[\alpha \Sigma^{AA} \pl_A \pl_A + d\omega(\pl)\right]d\Psi(y)=0\,,
\end{align}
i.e. all $y$s in the brackets are replaced by $\pl$, which contracts the indices of the fields in the brackets with $d\Psi$. Let us collect the two pieces of $d\omega$ together
\begin{align}
    d\omega(\Sigma)+d\omega(\bar{\Sigma})&= \tfrac12{\Sigma}^{CC}\left(\xi_{CB}y^B\xi_{CB}y^B 2(s-1)u^{s-2}+u^{s-1} \xi_{CC}\right)\chi_0^{(s)}+\notag\\
    &+\tfrac12\bar{\Sigma}^{C'C'}\left(\chi_1^{(s)} u^{s-2} y_{C'}y_{C'}+\chi_2^{(s)} u^{s-1} \xi_{C'C'} +\chi_3^{(s)} u^{s-2} \xi_{C'B}y^B\xi_{C'B}y^B \right)=\notag\\
    &= \tfrac12{\Sigma}^{CC}F_{CC}+\tfrac12\bar{\Sigma}^{C'C'}\bar{F}_{C'C'}\,.
\end{align}
Where all $y_A$s should be replaced with $\pl_A$. After some lengthy but straightforward calculation, see Appendix \ref{app:negativehelicity} for more detail, the contraction is found to be
\begin{align*}
    d\omega^{(n)}\wedge d\Psi^{(s)}&=\tfrac13 V^{-1/2} E^{AA'} \left( T^1 (\xi y)_{(A} y_{A')}+T^2 \epsilon_{AA'}\right)\,,
\end{align*}
where we defined (see Appendix \ref{app:negativehelicity} for the definitions of the coefficients $\acoef_{n,s}$)
\begin{align*}
    T^1&= u^{s-n} \frac{\acoef_{n,s}}{8 \rho (s-n+2)(s-n+1)} t^1\,,\\
    T^2&= u^{s-n+1} \frac{\acoef_{n-1,s}}{4 \rho} t^2\,,
\end{align*}
and (the sources have spin-$s$ with $\kappa_s$ and spin-$n$ with $\sigma_n$)
\begin{align*}
    t^1 r V^{3/2}&= -2V (n-1)n\sigma_n  \left(2 \rho \kappa_s'+(1+2s)\kappa_s\right)-4Vn \rho (s+1) \kappa_s \sigma_n'+2 n\rho (s+1) \kappa_s \sigma_n V'\,,\\
    t^2 V^{3/2}&= 2 (n-1) V \left(n \sigma_n \left(2 \rho \kappa_s'+(1+2s)\kappa_s
    \right)+2 \rho (s+1) \kappa_s \sigma_n'\right)+\\
    &-2 \rho \sigma_n V' \left(4 n \rho \kappa_s'+(n (s-1)+s+1) \kappa_s\right)\,.
\end{align*}
The last ingredient of the equation is the free part
\begin{align}\label{dPsicomputed}
    \alpha \Sigma^{AA} \pl_A \pl_A d\Psi(y)&= -\tfrac23 \alpha V^{-1/2} E^{AB'} (K^1 (\xi y)_{(A}y_{B')} + K^2\epsilon_{AB'})\,,\\
    K^1&= u^{s-2} 2s(s-1)[(2s+1) \kappa_s +2 \rho \kappa'_s]\notag\,,\\
    K^2&= u^{s-1} s^2[(2s+1) \kappa_s + 2 \rho\kappa'_s]\notag\,.
\end{align}
Let us factor out the free solutions and consider the following ansatz
\begin{align}\label{factoringout}
    \kappa_{s}&\rightarrow \rho ^{-1/2-s} {\kappa}_s[V]\,,& 
    \sigma_{s}&\rightarrow  V^{-1/2} {\sigma}_s[V]\,. 
\end{align}
Upon dividing the equations by a certain numerical factor to give the same unit normalization to the free term, we get
\begin{align}\label{psitwoeq}
\begin{aligned}
    \kappa_{s'}'+\frac{g_{s,n}}{\alpha}   
\left\{
(n-1)V^{-1}\sigma_n\kappa'_s
-(s+1)\left[V^{-2}\sigma_n -V^{-1}\sigma'_n\right]\kappa_s
\right\}=0\,,\\
    \kappa_{s'}'-\frac{g_{s,n}}{\alpha}
\left\{
(nV^{-1}-v_0 V^{-2})\sigma_n 
\kappa'_s
+(s+1)\left[V^{-2}\sigma_n+\frac{n-1}{n}V^{-1}\sigma'_n\right]\kappa_s
\right\}=0\,,
\end{aligned}
\end{align}
where we defined 
\begin{align}
    g_{s,n}&=4^{ n-2}n\left[\frac{s!}{(s-n+2)!}\right]^2\,.
\end{align}
The spins of the source and of the output are related by $s'=s-n+2$ or, in other words, $\hat{s'}=\hat{s}-\hat{n}$. It appears problematic to have two different equations for just one functions $\kappa_s$. Nevertheless, as the initial data let us take the free solutions for $\kappa_s$ and $\sigma_n$, i.e. $\kappa_s=b_{s,0}$ and $\sigma_n=c_{n,0}$ (with the free solutions factored out in \eqref{factoringout}, the initial data are just constants). It is already a miracle that both tensor structures lead to the same equation, which is solved by 
\begin{align}
    \kappa_{s'}&=-\frac{(s+1) g_{s,n}}{\alpha  V }b_{s,0}c_{n,0}\,.
\end{align}
Therefore, the equations are not overdetermined at the first order in perturbation theory. We could also see that for $\Psi_s$-perturbations the same pattern persists as for $\omega$, i.e. the solutions admit an expansion in the negative powers of $V$ and the deeper we go into the perturbation theory the higher the powers one meets. Therefore, it is sufficient to evaluate the general response to $\sigma_n= c_{n,m} V^{-m}$ and $\kappa_s=b_{s,k} V^{-k}$, i.e. for monomials. Since the two equations \eqref{psitwoeq} are supposed to be consistent, it is easier to solve the first one, which gives
\begin{align}
    \kappa_{s-n+2}\ni -\frac{ V^{-k-m-1} g_{s,n} (k (n-1)+(m+1) (s+1))}{\alpha  (k+m+1)}c_{n,m}b_{s,k}\,.
\end{align}
Similarly to the $\omega$ perturbation theory, we can formalize this by defining an action of the commutative non-associative algebra with basis $V^m_j$ on generators $U_{s}^{k}$ via (we use hatted variables $\hat{s}$ below)
\begin{align}
    U_{s}^{k} \lhd V_{j}^{m} &=-\frac{  g_{s+2,j+2} (k (j+1)+(m+1) (s+3))}{\alpha  (k+m+1) } U^{k+m+1}_{s-j}
\end{align}
Now let us have a look at the second order consistency. We need to consider various contributions of the form
\begin{align}\label{quasiass}
    (\Psi_{s+s_1+s_2-4} \lhd \omega_{s_1})\lhd \omega_{s_2}+(\Psi_{s+s_1+s_2-4} \lhd \omega_{s_2})\lhd \omega_{s_1} + 2 \Psi_{s+s_1+s_2-4}\lhd (\omega_{s_1}\circ \omega_{s_2})\,.
\end{align}
The terms in the brackets create intermediate fields that are used at the next order and the $\circ$/$\lhd$-operations show in which order we need to merge the fields. Eventually, we form a source for the spin-$s$ field $\Psi_s$. It is then can be checked with the formulas above that the two equations \eqref{psitwoeq} lead to the same $\Psi_s$. This supports, but does not prove, the statement that various contributions at the $n$th order conspire to give the same solution to the two equations \eqref{psitwoeq}. We leave the higher order consistency as a conjecture. Note that \eqref{quasiass} is not a form of ``associativity'' as it involves very specific generators: the $V_{s_1}^0$, $V_{s_2}^0$ and $U^0_{s+s_1+s_2-4}$, i.e. those that represent free solutions. The consistency condition at higher orders would be the statement that the sum of all possible terms of type
\begin{align}
    \sum_{\text{bracketings }} \Psi \lhd \omega\circ ... \omega \lhd ...\omega \circ ...\circ \omega 
\end{align}
gives the same solution for the two equations \eqref{psitwoeq} if we start with the free solutions for $\Psi$ and $\omega$s.

\section{Conclusions and Discussion}
\label{sec:conclusions}

In this paper we continue to explore the solution space of chiral higher-spin gravity and its two contractions: HS-SDYM and HS-SDGR. One interesting feature of HS-SDYM and HS-SDGR models studied in the paper is the pyramid structure of interactions and gauge transformations. For example, the spin-two forms a closed subsector of HS-SDGR. Usually, higher-spin symmetries make all spins contribute to any give one. In particular, one can try to redefine any given spin with the help of a gauge transformation. For example, one can map a black hole into something that does not look like a black hole anymore in its spin-two subsector, see \cite{Ammon:2011nk} for an explicit example in a $3d$ matter-free higher-spin theory (Chern-Simons theory). This is impossible in HS-SDYM and HS-SDGR. Therefore, in these theories it should be possible to use the usual geometrical concepts, e.g. horizon. In general, one has to develop new higher-spin symmetry friendly tools to replace the usual geometric notions, see e.g. \cite{Sezgin:2011hq}. For example, it was shown in \cite{Ivanovskiy:2025kok,Ivanovskiy:2025ial} that the chiral higher-spin gravity as well as HS-SDGR cannot couple to a point particle.

There is an obvious extension of the present work: one can uplift the higher-spin Taub-NUT solutions of HS-SDYM an HS-SDGR to the full chiral theory in flat space and then to anti-de Sitter space.\footnote{In this regard, some of the techniques and ideas from \cite{Iazeolla:2007wt} can be very much relevant since it has self-dual solutions of Type-D, see also \cite{Iazeolla:2011cb}. However, the solutions are not in a gauge that could potentially lead to local interactions of the chiral theory or its contractions: HS-SDYM and HS-SDGR. One can also rephrase the problem of finding the Taub-NUT solution as a problem of constructing a field-redefinition that would make the interactions chiral and local in the gauge used in \cite{Iazeolla:2007wt}.  } Apart from the general interest in the space of higher-spin instantons, the AdS higher-spin Taub-NUT solution should play an important role in the chiral higher-spin gravity/Chern-Simons vector model duality \cite{Skvortsov:2018uru,Sharapov:2022awp,Jain:2024bza,Aharony:2024nqs}, which is an extrapolation of the ideas of \cite{David:2020fea} for the vector model/higher-spin duality \cite{Klebanov:2002ja, Sezgin:2002rt,Sezgin:2002rt, Leigh:2003gk, Giombi:2011kc}. In this case, the Taub-NUT solution should be dual the multiplet of higher-spin currents that has, say, only $\langle J_{+s} J_{+s}\rangle $ two-point functions nonvanishing, while $\langle J_{-s} J_{-s}\rangle =0$.

The recent advances in the twistor description of the chiral higher-spin gravity and its contractions \cite{Mason:2025pbz, Tran:2025uad} should pave the way to direct applications of twistor techniques to constructing exact solutions, see e.g. \cite{Bu:2022iak,Bittleston:2023bzp,Adamo:2023fbj,Adamo:2025fqt} for closely related results.

\section*{Acknowledgments}
\label{sec:Aknowledgements}
We are grateful to Carlo Iazeolla and Kirill Krasnov for many stimulating discussions. This project has received funding from the European Research Council (ERC) under the European Union’s Horizon 2020 research and innovation programme (grant agreement No 101002551).

\begin{appendix}

\section{Normalization}
\label{app:normalization}
We want to adjust the coupling constants in the action to make them consistent with the equations. In the action, one can rescale (fluctuations) fields $\omega_s \rightarrow \alpha_s \omega_s$ without affecting any physics, the net result being
\begin{align}
    \Sigma^{AA}\wedge d\omega^{A(2s)}+\tfrac12 \sum_{s_1+s_2-2=s}\frac{a_{s_1,s_2}\alpha_{s_1}\alpha_{s_2}}{a_{2,s}\alpha_s}\, d\omega^{A(2s_1-2)}\wedge d\omega^{A(2s_2-2)} &=0 
\end{align}
The relevant part of the equations of motion is
\begin{align}
    d\omega &= \pl^C\omega \wedge \pl_C \omega+...
\end{align}
This is not a complete equation of motion, but we only need to look at the $\bry=0$ components and the one linear in $\bry$. We can make the following ansatz
\begin{align}
    \omega&= -\tfrac12 e^{AA'} y_A \bry_{A'} -\tfrac14 \omega^{AA} y_A y_A + \sum_s \beta_s \omega_{A(2s-2)} \, y^A...y^A +\sum_s \omega_{A(2s-3),A'} \, y^A...y^A\, \bry^{A'}
\end{align}
With the normalization as above, we have $d\omega^{AA}=\Sigma^{AA}$ and at the free level
\begin{align}
    \beta_s d\omega^{A(n)}&= e\fud{A}{B'} \omega^{A(n-1),B'}
\end{align}
This allows us to evaluate the interaction term
\begin{align}
    d\omega^{A(n)}\wedge d\omega^{A(m)}&= -\tfrac12 \Sigma^{AA}\omega\fud{A(n-1),B'}\wedge \omega^{A(m-1),B'}
\end{align}
The equations of motion, as obtained from the equation acquire the following form 
\begin{align}
    \Sigma^{AA}\wedge\left( d\omega^{A(2s)}+\tfrac12(-\tfrac12) \sum_{s_1+s_2-2=s}\frac{a_{s_1,s_2}\alpha_{s_1}\alpha_{s_2}}{a_{2,s}\alpha_s\beta_{s_1}\beta_{s_2}}\, \omega\fud{A(n-1),B'}\wedge \omega^{A(m-1),B'} \right)&=0 
\end{align}
The equations of motion, as obtained from the chiral theory, need the source term
\begin{align}
    \pl^C\omega \wedge \pl_C \omega+...\ni -\omega\fud{A(n-1),B'}\wedge \omega^{A(m-1),B'} \, y^A...y^A
\end{align}
Finally, we get
\begin{align}
    \Sigma^{AA}\wedge \left(d\omega^{A(2s)}+\tfrac{1}{\beta_s} \sum_{s_1+s_2-2=s}\omega\fud{A(2s_1-2),B'}\wedge \omega^{A(2s_2-2),B'}\right)=0
\end{align}
Therefore, to relate the normalizations we have to impose
\begin{align}
    -\frac14\frac{a_{s_1,s_2}\alpha_{s_1}\alpha_{s_2}}{a_{2,s}\alpha_s\beta_{s_1}\beta_{s_2}}=\frac{1}{\beta_s}
\end{align}
We choose $\alpha_s=1$ (not to rescale anything in the action), $\beta_s=-1/4$ and $a_{s_1,s_2}=a_{2,s}=1$.

For HS-SDYM it is easier to normalize the action since the relevant part of the equations is 
\begin{align}
    d\omega &= \omega \wedge \omega+...
\end{align}
where $\omega$ is assumed to carry some color, which makes the r.h.s. nonvanishing. Therefore, in terms of generating functions, we have $F=d\omega-\omega\wedge \omega$. 

\section{Solution to HS-SDGR}
\subsection{Positive helicity}
\label{app:positivehelicity}
The differential equation for $\sigma _{s}$ (note that we convert the equation by making $\sigma $ depend on $V$
instead of $\rho $)%
\begin{equation}
\sigma _{s}\left( V\right) +2V\sigma _{s}^{\prime }\left( V\right) =\sum
_{\substack{ s_{1}+s_{2}=s+2  \\ s_{1},s_{2}\geq 3}}f_{s_{1},s_{2}}\left(
V\right) \ ,  \label{eqtosolve}
\end{equation}%
where
\begin{align}
f_{s_{1},s_{2}}\left( V\right) &=\frac{1}{2s-1}\Big\{ \left[ \left(
s+1\right) V^{\frac{1}{2}}-s_{1}v_{0}V^{-\frac{1}{2}}\right] s_{2}\sigma
_{s_{1}}^{\prime }\left( V\right) \sigma _{s_{2}}\left( V\right) +\\
&\quad+\left[
\left( s+1\right) V^{\frac{1}{2}}-s_{2}v_{0}V^{-\frac{1}{2}}\right]
s_{1}\sigma _{s_{1}}\left( V\right) \sigma _{s_{2}}^{\prime }\left( V\right)+
\notag \\
&\quad +\left[ \left( s+1\right) \left( \frac{1}{2}s-2s_{1}s_{2}+1\right)
V^{-\frac{1}{2}}+s_{1}s_{2}v_{0}V^{-\frac{3}{2}}\right] \sigma
_{s_{1}}\left( V\right) \sigma _{s_{2}}\left( V\right) \Big\} \ .
\end{align}%
The solution to (\ref{eqtosolve}) for $\sigma _{s}$ is%
\begin{equation}
\sigma _{s}=\tfrac{1}{2}V^{-\frac{1}{2}}\int V^{-\frac{1}{2}}\sum_{\substack{ %
s_{1}+s_{2}=s+2  \\ s_{1},s_{2}\geq 3}}f_{s_{1},s_{2}}\left( V\right) \ dV\ ,
\label{sigsint}
\end{equation}%
where the indefinite integral gives an integration constant, which
multiplied by $\frac{1}{2}V^{-\frac{1}{2}}$ gives the free part of the
solution.

\subsection{Negative helicity}
\label{app:negativehelicity}
In order to contract $d\omega(\pl)$ with $d\Psi(y)$ we introduce $\Delta=(\pl\xi \pl)$, which is the dual of $u=(y\xi y)$. One needs the following contractions
{\allowdisplaybreaks
\begin{align*}
   \Delta^n (u^s)&=\strA{n,s}=\acoef_{n,s} u^{s-n} \,,\\
   \Delta^n (u^{s} y_A y_A)&=\strB{n,s}_{AA}=b^1_{n,s} u^{s-n}y_Ay_A+b^2_{n,s} u^{s-n+1}\xi_{AA}\,,\\
   \Delta^{n}\pl_C\pl_C( u^s)&=\strC{n,s}_{CC}= c^1_{n,s}u^{s-n-1}\xi_{CC}+c^2_{n,s} u^{s-n-2} (\xi y)_C (\xi y)_C\,,\\
    \Delta^{n}(\xi\pl)_C(\xi\pl)_C( u^s)&=\strCxi{n,s}_{CC}= c^1_{n,s}\rho u^{s-n-1}\xi_{CC}+c^2_{n,s}\rho^2 u^{s-n-2} y_C y_C\,,\\
   \Delta^{n}\pl_C\pl_C( u^s y_A y_A)&=\strD{n,s}_{CC,AA}= d^1_{n,s}u^{s-n-2}(\xi y)_C(\xi y)_Cy_A y_A+d^2_{n,s}u^{s-n-1}\xi_{CC}y_A y_A +\\
   &+d^3_{n,s}u^{s-n-1}(\xi y)_C \epsilon_{CA}y_A+
   +d^4_{n,s}u^{s-n}\epsilon_{CA}\epsilon_{CA}+\\
   &+d^5_{n,s}u^{s-n-1}(\xi y)_C (\xi y)_C \xi_{AA}+d^6_{n,s}u^{s-n}\xi_{CC}\xi_{AA}\,,\\
   \Delta^{n}(\xi\pl)_C(\xi\pl)_C( u^s y_A y_A)&=\strDxi{n,s}_{CC,AA}= d^1_{n,s}u^{s-n-2}\rho^2 y_Cy_Cy_Ay_A+d^2_{n,s}u^{s-n-1}\rho \xi_{CC}y_A y_A +\\
   &+d^3_{n,s}u^{s-n-1}\rho y_C \xi_{CA}y_A+
   +d^4_{n,s}u^{s-n}\xi_{CA}\xi_{CA}+\\&+d^5_{n,s}u^{s-n-1}\rho^2y_Cy_C \xi_{AA}+d^6_{n,s}u^{s-n}\rho\xi_{CC}\xi_{AA}\,,
\end{align*}}
\noindent where
{\allowdisplaybreaks
\begin{align}
    \acoef_{n,s}&= 4^n[s!/(s-n)!]^2\rho^n\,,\\
    b^1_{n,s}&= \acoef_{n,s}+8 \rho n(s-n+1)\acoef_{n-1,s} +16 n(n-1) \acoef_{n-2,s} (s-n+2)(s-n+1)\rho^2\,,\\
    b^2_{n,s}&=2n \acoef_{n-1,s} +8n(n-1)\acoef_{n-2,s}\rho (s-n+2)\,,\\
    c^1_{n,s}&=2(s-n)\acoef_{n,s}\,,\\
    c^2_{n,s}&=4(s-n)(s-n-1)\acoef_{n,s}\,,\\
    d^1_{n,s}&=b^1_{n,s}4(s-n)(s-n-1)\,,\\
    d^2_{n,s}&=b^1_{n,s}2(s-n)\,,\\
    d^3_{n,s}&=b^1_{n,s}8(s-n)\,,\\
    d^4_{n,s}&=b^1_{n,s}2\,,\\
    d^5_{n,s}&=b^2_{n,s}4(s-n)(s-n+1)\,,\\
    d^6_{n,s}&=b^2_{n,s}2(s-n+1)\,.
\end{align}}\\
\noindent In fact, $b^1_{n,s}=\frac{(s+1)(s+2)}{(s-n+1)(s-n+2)}\acoef_{n,s}$ and $b^2_{n,s}=2n \frac{s+1}{s-n+2}\acoef_{n-1,s}$, but above we have revealed how different coefficients are related to each other. Since the equation is a three-form, it is necessary to introduce the basis of those
\begin{align}
    E^{AA'}&\equiv e\fdu{B}{A'}\wedge \Sigma^{AB}\,, \\
    e^{AA'}
    \wedge \Sigma^{BB}&=+\tfrac23 \epsilon^{AB} E^{BA'}\,,\\
     e^{AA'}
    \wedge \Sigmab^{B'B'}&=-\tfrac23 \epsilon^{A'B'} E^{AB'}   \,,
\end{align}
together with some useful identities. Now we need to compute
\begin{align}
    d\omega(\pl)\wedge d\Psi(y)&= \tfrac 12 e^{AA'}\Sigma_{CC} F_{CC} \nabla_{AA'}\Psi+\tfrac 12 e^{AA'}\Sigmab_{C'C'} \bar{F}_{C'C'} \nabla_{AA'}\Psi=\\
    &-\tfrac 13 E^{CA'} F_{MC}\nabla\fud{M}{A'}\Psi +\tfrac 13 E^{AC'}F_{M'C'}\nabla\fdu{A}{M'}\Psi\,.
\end{align}
We can put together $d\omega$ of spin $n$ and $d\Psi$ of spin $s$ to find for the coefficients of the two terms above
\begin{align*}
    -\tfrac13 V^{-1/2}E^{CA'}:\,& 
    2(n-1) \chi_0^n s\kappa_s\times \strDxi{n-2,s-1}\fdu{CM|A'}{M} + 
    2(n-1) \chi_0^n \kappa'_s\times \strCxi{n-2,s}_{CM}\xi\fud{M}{A'}+\\
    &\chi_0^n s\kappa_s \times\xi_{CM}\strB{n-1,s-1}\fud{M}{A'}+
    \chi_0^n \kappa'_s\times \strA{n-1,s} \rho \epsilon_{CA'}
\end{align*}
\begin{align*}
    +\tfrac13 V^{-1/2}E^{AC'}:\,& 
    \chi_1^n s\kappa_s \times \strD{n-2,s-1}\fdu{C'M'|A}{M'}+
    \chi_1^n \kappa'_s \times \strC{n-2,s}_{C'M'} \xi\fdu{A}{M'}+ \\
    &\chi_2^n s\kappa_s\times \xi_{C'M'}\strB{n-1,s-1}\fdu{A}{M'}+
    \chi_2^n \kappa'_s \times \strA{n-1,s} \rho \epsilon_{C'A}+\\
    &\chi_3^n s\kappa_s\times \strDxi{n-2,s-1}\fdu{C'M'|A}{M'}+
    \chi_3^n \kappa'_s\times \strCxi{n-2,s}_{C'M'}\xi\fdu{A}{M'}\,,
\end{align*}
where some of the tensor structures are in fact the same and we need just
\begin{align*}
    \xi_{C'M'}\strB{n,s}\fdu{A}{M'}&= b^1_{n,s} u^{s-n} (\xi y)_{C'} y_A +b^2_{n,s} u^{s-n+1}\epsilon_{C'A}\rho\,,\\
    \strC{n,s}_{CM}\xi\fud{M}{A'}&= c^1_{n,s} u^{s-n-1} \epsilon_{CA'}\rho+c^2_{n,s}u^{s-n-2} (\xi y)_C \rho y_{A'}\,,\\
    \strCxi{n,s}_{C'M'}\xi\fdu{A}{M'}&=c^1_{n,s} u^{s-n-1} \epsilon_{C'A}\rho^2+c^2_{n,s}u^{s-n-2} (-\rho^2)(\xi y)_{A}  y_{C'}\,,\\
    \strD{n,s}\fdu{CM'|A}{M'}&=d^1_{n,s}u^{s-n-1}(\xi y)_C y_A+d^2_{n,s}u^{s-n-1}(\xi y)_{C} y_A +\\
   &+d^3_{n,s}u^{s-n-1}((\xi y)_C y_A+\tfrac14 u \epsilon_{CA})
   +d^4_{n,s}u^{s-n}\tfrac32\epsilon_{CA}+\\
   &+d^5_{n,s}u^{s-n-1}\rho (\xi y)_C y_{A}+d^6_{n,s}u^{s-n}\rho\epsilon_{CA}\,,\\
   \strDxi{n,s}\fdu{CM|A}{M}&=d^2_{n,s}u^{s-n-1}\rho (\xi y)_C y_A +d^3_{n,s} u^{s-n} (-\rho\tfrac14) \epsilon_{CA}
   +d^4_{n,s}u^{s-n}\rho (-\tfrac12)\epsilon_{CA}+\\&+d^5_{n,s}u^{s-n-1}(-)\rho^2(\xi y)_Ay_C +d^6_{n,s}u^{s-n}\rho^2\epsilon_{CA}\,.
\end{align*}
At the end we can write everything as
\begin{align*}
    d\omega^{(n)}\wedge d\Psi^{(s)}&=\tfrac13 V^{-1/2} E^{AA'} \left( T^1 (\xi y)_{(A} y_{A')}+T^2 \epsilon_{AA'}\right)\,,
\end{align*}
where two irreducible tensor structures are isolated. 
Once we plug in the actual coefficients we find
\begin{align*}
    T^1&= u^{s-n} \frac{\acoef_{n,s}}{8 \rho (s-n+2)(s-n+1)} \Big( \chi_1^n [2 \kappa'_s+\tfrac{(2 s+1) }{\rho}\kappa_s] +\chi_2^n 2(s+1) \kappa_s +\chi_3^n [\kappa_s-2 \rho \kappa'_s]+\\
    &\qquad + \chi_0(4 (n-1) \rho \kappa'_s-2 (n+s) \kappa_s)\Big)\,,\\
    T^2&= u^{s-n+1} \frac{\acoef_{n-1,s}}{4 \rho} \Big( \chi_1^n [-2\rho \kappa'_s-(1+2s)\kappa_s] -2\rho\chi_2^n [s\kappa_s+2 \rho \kappa'_s]+\rho\chi_3^n[\kappa_s-2 \rho \kappa'_s]+\\
    &\qquad + \chi_0 (-4n\rho^2 \kappa'_s-2(s-n+1)\rho \kappa_s)\Big)\,.
\end{align*}
Now we can also recall what $\chi_{0,1,2,3}$ are and plug them into the expressions above to get the rest of the formulas in the main text.

\section{HS-SDYM negative helicity}
\label{app:SDYMNegative}
The equation we want to solve is:%
\begin{equation}
\Sigma ^{AA}\wedge \partial _{A}\partial _{A}\left( \alpha d\Psi
^{(s)}-\sum_{q}\left[ \omega ^{(q)}(\partial ),\Psi ^{(s+q-1)}\right] \right)
=0\ ,
\end{equation}%
and let us investigate the term in the sum:%
\begin{equation}
\Sigma ^{AA}\wedge \partial _{A}\partial _{A}\omega ^{(q)}(\partial )\Psi
^{(s+q-1)}
\end{equation}%
by substituting the ansatz%
\begin{equation*}
\omega ^{(q)}=e^{CC^{\prime }}\partial _{C}\bar{\partial}_{C^{\prime }}\left[
u^{q-1}w\sigma ^{(q)}\right] \text{ \ \ and \ \ }\Psi ^{(s)}=u^{s}\kappa
^{(s)}\ .
\end{equation*}
First, $\omega ^{(q)}$ can be written as%
\begin{equation}
\omega ^{(q)}=e^{CC^{\prime }}\left[ 2\left( q-1\right) u^{q-2}\left( \xi
y\right) _{C}y_{C^{\prime }}+u^{q-1}\epsilon _{CC^{\prime }}\right] \sigma
^{(q)}\ ,
\end{equation}%
then $\omega ^{(q)}(\partial )\Psi ^{(s+q-1)}$ can be calculated:%
\begin{align}
\omega ^{(q)}&(\partial )\Psi ^{(s+q-1)}=e^{CC^{\prime }}\left[ 2\left( q-1\right) \Delta ^{q-2}\left( \xi
\partial \right) _{C}\partial _{C^{\prime }}+\Delta ^{q-1}\epsilon
_{CC^{\prime }}\right] u^{s+q-1}\sigma ^{(q)}\kappa ^{(s+q-1)} \notag \\
&=e^{CC^{\prime }}\left[ 2\left( q-1\right) c_{q-2,s+q-1}^{1}\rho
u^{s}\epsilon _{CC^{\prime }}-2\left( q-1\right) c_{q-2,s+q-1}^{2}\rho
u^{s-1}y_{C}\left( \xi y\right) _{C^{\prime }}+\acoef_{q-1,s+q-1}u^{s}\epsilon
_{CC^{\prime }}\right] \sigma ^{(q)}\kappa ^{(s+q-1)}  \notag \\
&=e^{CC^{\prime }}\left[ 2\left( s+q\right) c_{q-2,s+q-1}^{1}\rho
u^{s}\epsilon _{CC^{\prime }}-2\left( q-1\right) c_{q-2,s+q-1}^{2}\rho
u^{s-1}y_{C}\left( \xi y\right) _{C^{\prime }}\right] \sigma ^{(q)}\kappa
^{(s+q-1)}\ .  \notag 
\end{align}%
Next, using $\Sigma ^{AA}\wedge e^{CC^{\prime }}=\frac{2}{3}\epsilon
^{CA}E^{AC^{\prime }}$ we further calculate%
\begin{align}\notag
\Sigma ^{AA}\wedge \partial _{A}&\partial _{A}\omega ^{(q)}(\partial )\Psi
^{(s+q-1)} 
=\\
&=\frac{2}{3}E^{CC^{\prime }}\sigma ^{(q)}\kappa ^{(s+q-1)}  \notag 
\left\{ 2\left( s+q\right) c_{q-2,s+q-1}^{1}\rho \left[ 2su^{s-1}\xi
_{CC^{\prime }}+4s\left( s-1\right) u^{s-2}\left( \xi y\right) _{C}\left(
\xi y\right) _{C^{\prime }}\right] \right.   \notag \\
&\left. +2\left( 2s+1\right) \left( q-1\right) c_{q-2,s+q-1}^{2}\rho \left[
u^{s-1}\xi _{CC^{\prime }}+2\left( s-1\right) u^{s-2}\left( \xi y\right)
_{C}\left( \xi y\right) _{C^{\prime }}\right] \right\}   \notag \\
&=\frac{2}{3}E^{CC^{\prime }}\sigma ^{(q)}\kappa ^{(s+q-1)}\left(
2q-1\right)   \left\{ c_{q-1,s+q-1}^{1}u^{s-1}\xi _{CC^{\prime
}}+c_{q-1,s+q-1}^{2}u^{s-2}\left( \xi y\right) _{C}\left( \xi y\right)
_{C^{\prime }}\right\} \ .\notag
\end{align}

\end{appendix}

\footnotesize
\providecommand{\href}[2]{#2}\begingroup\raggedright\endgroup

\end{document}